\documentclass[aps,prl,article,twocolumn,showpacs,preprintnumbers,amsmath,amssymb,superscriptaddress]{revtex4}
\date{\today}
\usepackage{epsfig}
\usepackage{subfigure}
\usepackage{graphicx}
\usepackage{dcolumn}
\usepackage{bm}
\usepackage[colorlinks,linkcolor=blue,hyperindex,CJKbookmarks]{hyperref}
\usepackage{float}
\usepackage{hyperref}
\usepackage{comment}
\usepackage{color}
\newcommand{\kw}[1]{\textcolor{black}{#1}}
\newcommand{\md}[1]{\textcolor{black}{#1}}

\begin{document}

\title{Jahn-Teller effect in systems with strong on-site spin-orbit coupling}

\author{Ekaterina M. Plotnikova}
\affiliation{IFW Dresden, Helmholtzstr. 20, 01069 Dresden, Germany}
\email[]{e.plotnikova@ifw-dresden.de}

\author{Maria Daghofer}
 \affiliation{Institute for Functional Materials and Quantum Technologies, University of Stuttgart,
Pfaffenwaldring 57
D-70550 Stuttgart, Germany}

\author{Jeroen van den Brink}
 \affiliation{IFW Dresden, Helmholtzstr. 20, 01069 Dresden, Germany}

\author{Krzysztof Wohlfeld}
\affiliation{Stanford University and SLAC National Accelerator Laboratory, 2575 Sand Hill Rd, Menlo Park, CA 94025 USA}
\affiliation{Institute of Theoretical Physics, Faculty of Physics, University of Warsaw, Pasteura 5, PL-02093 Warsaw, Poland}

\date{\today}
\begin{abstract}
When strong spin-orbit coupling removes orbital degeneracy, it would at the same time appear to render the Jahn-Teller mechanism ineffective. We discuss such a situation, the $t_{2g}$ manifold of iridates, and show that, while the Jahn-Teller effect does indeed not affect the $j_{\textrm{eff}}=1/2$ antiferromagnetically ordered ground state, it leads to distinctive signatures in the  $j_{\textrm{eff}}=3/2$ spin-orbit exciton. It allows for a hopping of the spin-orbit exciton between the nearest neighbor sites without producing defects in the $j_{\textrm{eff}}=1/2$  antiferromagnet. This arises because the lattice-driven Jahn-Teller mechanism only couples to the orbital degree of freedom, but is not sensitive to the phase of the wave function that defines isospin $j_z$. This contrasts sharply with purely electronic propagation, which conserves isospin, and presence of Jahn-Teller coupling can explain some of the peculiar features of  measured resonant inelastic x-ray scattering spectra of Sr$_2$IrO$_4$.
\end{abstract}
\maketitle

{\it Introduction} The discovery that spin-orbit
coupling (SOC) can induce bulk
insulators with conducting edge states, which  are symmetry protected against back
scattering, has in recent years revived interest in spin-orbit coupled
materials~\cite{Hasan2010, XiaoLiang2011}. While typical topological insulators
are at most weakly correlated, the interplay of electron-electron
interaction and spin-orbit coupling has also received enhanced attention:
On one hand, the combination was soon discovered as a promising route to alternative topologically
nontrivial states, from topological Mott~\cite{Pesin:2010ju,WitczakKrempa2014} over fractional Chern~\cite{doi:10.1142/S021797921330017X}
insulators to a potential
realization~\cite{Jackeli2009,Chaloupka2010} for  Kitaev's celebrated spin-liquid phase with its
anyonic excitations~\cite{Kitaev2006, Baskaran2007}. On the other
hand, spin-orbit coupled and correlated square-lattice iridates are
emerging as a sister-system to high-$T_C$
cuprates~\cite{Kim2008,WangSenthil2011,Watanabe2013,PRL2012Kim,
  Naturecom2014Maria,Kim11072014,YKKim2015_dgap,2015arXiv150606557Y}. 

The cuprate-like physics and the Kitaev-Heisenberg model
supporting the spin liquid are both understood to arise as the low-energy
limit in iridium compounds like square-lattice Sr$_2$IrO$_4$~\cite{Kim2008,Jackeli2009,WangSenthil2011} and
honeycomb-lattice Na$_2$IrO$_3$~\cite{Chaloupka2010, Chaloupka2013}. In such iridates, the $t_{2g}$ levels of the $5d$
shell are almost filled, the single hole is subject to both strong SOC
and appreciable correlations. The $t_{2g}$ manifold can be described
as an effective angular momentum $l_{\textrm{eff}} =1$ and SOC locally 
couples spin ${\bf s}$ and ${\bf l}$ to a total angular momentum ${\bf j}$. The threefold orbital degeneracy of the $t_{2g}$ states is thus lifted by SOC and on-site
Hubbard interaction can subsequently open a charge gap and stabilize a localized
(pseudo)spin $j_{\textrm{eff}}=1/2$~\cite{Kim2008, Jackeli2009}. Due to the orbital
part of the $j_{\textrm{eff}}=1/2$ wave function, couplings between these effective
spins are sensitive to lattice geometry and support a variety of
quantum states. 

\begin{figure}[b]
\includegraphics[width=0.85\columnwidth]{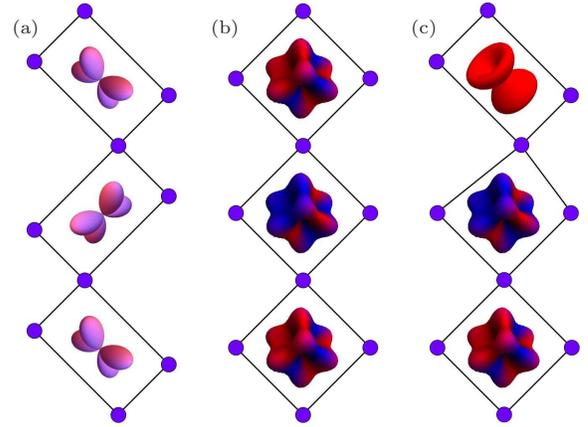}\llap{
  \parbox[b]{5.8in}{\scriptsize{(a)}\\\rule{0ex}{2.0in}
  }}
  \llap{
  \parbox[b]{3.8in}{\scriptsize{(b)}\\\rule{0ex}{2.0in}
  }}
   \llap{
  \parbox[b]{1.8in}{\scriptsize{(c)}\\\rule{0ex}{2.0in}
  }}
\caption{Cartoon picture showing the Jahn-Teller effect in systems {\it without} and {\it with} strong SOC:
(a) Weak SOC -- oxygen displacements following `conventional' Jahn-Teller effect for the ground state with e.g. the $d_{xz}/d_{yz}$ alternating orbital order.
(b) Strong SOC -- no oxygen displacements due to the quenched Jahn-Teller effect for the ground state with 
e.g. $|j_{\textrm{eff}}=1/2$, $j_z=1/2 \rangle / |j_{\textrm{eff}}=1/2$, $j_z=-1/2 \rangle$ alternating spin-orbital order (antiferromagnetic order of $j_{\textrm{eff}}=1/2$ isospins).
(c) Strong SOC -- oxygen displacements around the $|j_{\textrm{eff}}=3/2$, $j_z=-3/2 \rangle$ exciton (which `lives' in the antiferromagnetic $j_{\textrm{eff}}=1/2$ ground state) showing that such a system is Jahn-Teller active.
\label{fig:1}}
\end{figure}

A striking difference to $3d$ systems with negligible~\cite{Oles2005} SOC
is the lifting of the orbital degeneracy: a single hole (or electron)
in a $3d$ shell has an orbital degree of freedom in addition
to spin -- as opposed to the single $j_{\textrm{eff}}=1/2$ degree of freedom of the
$5d$ hole. As a consequence, an analogous $3d$ system can not only feature orbital order in
addition to magnetism, but the Jahn-Teller effect [cf.~Fig.~\ref{fig:1}(a)] would moreover be
expected to couple the orbital degree of freedom to the
lattice~\cite{Kanamori1960, Kugel1984, Nasu2013, Nasu2015}.  In
contrast, the quenching of the orbital degree of freedom by SOC removes the possibility
of orbital order and would at first sight also appear to suppress
Jahn-Teller effect and coupling to the lattice.  

In this Letter, we are nevertheless going to discuss the impact of the 
Jahn-Teller effect on $5d$ systems with strong SOC: While it is indeed
absent for the ground state consisting of  $j_{\textrm{eff}}=1/2$ pseudospins, see
Fig.~\ref{fig:1}(b), we are
going to show that it leaves clear signatures in the dynamics of
collective \emph{excitations} into the $j_{\textrm{eff}}=3/2$ sector
(i.e. excitons). 
\kw{As seen in Fig.~\ref{fig:1}(c), the Jahn-Teller
effect is here not quenched and can allow for 
a novel type of excitonic propagation. In particular, we propose
that the experimentally observed branch of the exciton dispersion with the minimum at the $\Gamma$ point~\cite{Naturecom2014Maria}, 
which can not be explained using superexchange alone, finds a natural
explanation within the present Jahn-Teller model.}  

{\it Finite Jahn-Teller for excited states} 
Since the SOC constant $\lambda>0$ is assumed
to be the largest energy scale involved, 
with $\lambda =0.382\;\textrm{eV}$ in Sr$_2$IrO$_4$~\cite{Naturecom2014Maria},
we start our analysis by diagonalizing this dominant term. This is
achieved by a basis change from ${\bf
  s}$ (the $s= 1/2$ spin)  and ${\bf l}$ (the effective $l_{\textrm{eff}}=1$
orbital moment) to total angular momentum ${\bf j}= {\bf l} + {\bf
  s}$. For a single hole in the $t_{2g}$ shell, SOC interaction
$\mathcal{H}_{\rm SOC}=\lambda \sum\limits_{i}{{\bf l}_i \cdot {\bf
    s}_i}$ becomes then $\mathcal{H}_{\rm SOC}= \lambda/2 \sum\limits_{i} ({\bf j}^2_j - {\bf
  s}^2_j - {\bf l}^2_j)$ and the ground state is given by the
doubly-degenerate $j_{\textrm{eff}}=1/2$ manifold, while the $j_{\textrm{eff}}=3/2$ manifold forms
the excited states at energy $3\lambda /2$. (A crystal-field splitting
$\Delta$ can explicitly be included into this analysis~\cite{Khaliullin2005,Jackeli2009}, but is
omitted here for clarity)

For $t_{2_g}$ electrons, the orbital operators ${\bf
  l }$ couple both to the tetragonal phonon modes $Q_2$ and $Q_3$ (the
$e_g$ modes) and to trigonal phonon modes 
$Q_4$, $Q_5$, and $Q_6$ (the $t_{2g}$ modes). After integrating out
the phonons, the Jahn-Teller interaction is expressed in terms of
${\bf l}$~\cite{Kugel1984}: 
\begin{align}
\label{HJT}
\mathcal{H}_{\rm JT}&=V \sum\limits_{\langle {\bf i}, {\bf j} \rangle}\left[\bigl(l_{\bf i}^z\bigr)^2-\frac{2}{3}\right]\left[\bigl(l_{\bf j}^z\bigr)^2-\frac{2}{3}\right] \nonumber \\
&+V \sum\limits_{\langle {\bf i},{\bf j} \rangle} 
\left[\bigl(l_{\bf i}^x\bigr)^2-\bigl(l_{\bf i}^y\bigr)^2\right] 
\left[\bigl(l_{\bf j}^x\bigr)^2-\bigl(l_{\bf j}^y\bigr)^2\right]
\nonumber \\
&+ \kappa V \sum \limits_{\langle {\bf i},{\bf j} \rangle}\left[\left(l_{\bf i}^xl_{\bf i}^y+l_{\bf i}^yl_{\bf i}^x\right)(l_{\bf j}^xl_{\bf j}^y+l_{\bf j}^yl_{\bf j}^x)+...\right].
\end{align}
The two classes of phonon modes lead to two {\it a priori}  independent
Jahn-Teller coupling constants $V_{e_g} \equiv V$ and  $V_{t_{2g}}\equiv \kappa V$; as
$V_{t_{2g}}$ is typically much smaller than $V_{e_g}$, we set
$\kappa=0.1$. The Jahn-Teller interaction scale $V$ can  from
experiment~\cite{Gretarsson2015} be inferred to be non-negligible, but
as its strength is at present unclear, we leave it as a free
parameter.  

The Jahn-Teller term $H_{\rm JT}$ is now, via straightforward but
tedious calculations, transformed into the eigenbasis of
$\mathcal{H}_{\rm SOC}$, i.e., written in terms of $j$ states: 
\begin{align}
\mathcal{H}_{\rm JT}\! =\! \mathcal{H}_{\rm JT} (1/2, 1/2)\! +\! \mathcal{H}_{\rm JT}(3/2, 1/2)\!+\! \mathcal{H}_{\rm JT}(3/2, 3/2).
\end{align}
The first term $\mathcal{H}_{\rm JT}(1/2, 1/2)$ denotes the
Jahn-Teller interaction between two $j_{\textrm{eff}}=1/2$ states -- it vanishes as
expected, reflecting the quenching of orbital physics within the
$j_{\textrm{eff}}=1/2$ subshell. The last term $\mathcal{H}_{\rm JT}(3/2, 3/2)$
between two $j_{\textrm{eff}}=3/2$ states can only contribute if a large number of
$j_{\textrm{eff}}=3/2$ states are present  and is thus strongly suppressed at large $\lambda$. The term $\mathcal{H}_{\rm
  JT}(3/2, 1/2)$ describes the interaction between one $j_{\textrm{eff}}=1/2$ and one
$j_{\textrm{eff}}=3/2$ site: Even at strong SOC,
this term becomes relevant when an (iso)orbital excitation raises a
single hole into a $j_{\textrm{eff}}=3/2$ state~\cite{PRL2012Kim, Naturecom2014Maria}. 

\kw{{\it Model}}
The $j_{\textrm{eff}}=3/2$ excitation, an exciton, can be created in resonant inelastic
X-ray scattering (RIXS) and has been discussed in two recent theoretical and
experimental studies~\cite{PRL2012Kim, Naturecom2014Maria}. 
It is described by the Green function 
\begin{flalign}
\label{green}
G({\bf k},\omega)=Tr {\langle 0 | \hat{\chi}_{ {\bf k}}\frac{1}{\omega-{H}+i\delta}
\hat{\chi}_{ {\bf k}}^\dagger| 0 \rangle}, 
\end{flalign}
where the $\hat{\chi}_{{\bf k}}^\dagger$ is a vector of four creation
operators that create an exciton with momentum ${\bf k}$ and isospin
quantum number $j_z = \pm 1/2, \pm 3/2$.
The Hamiltonian $H$ describes the dynamics of the exciton coupling to a
background of $j_{\textrm{eff}}=1/2$ isospins, a minimal Hamiltonian is
\begin{align}
	\label{H_exc}
	    {H} = {H}^{\textrm{mag}}_{\rm SE}+  {H}^{\textrm{exc}}_{\rm SE} + {H}^{\textrm{exc}}_{\rm JT}\;. 
\end{align}
The first term ${H}^{\textrm{mag}}_{\rm SE}$ is the superexchange interaction between $j_{\textrm{eff}}=1/2$
isospins, where we include up to third-neighbor processes $\propto \{J_1, J_2, J_3\} $ [see Eq.~(6) of the supplemental materials of 
Ref.~\cite{Naturecom2014Maria}]. It stabilizes an alternating order of
$j_{\textrm{eff}}=1/2$ isospins with magnon-like excitations~\cite{PRL2012Kim, Naturecom2014Maria}. The terms ${H}^{\textrm{exc}}_{\rm
  SE}$ and ${H}^{\textrm{exc}}_{\rm JT}=\mathcal{H}_{\rm JT}(3/2, 1/2)$
describe superexchange and Jahn-Teller interaction between one
$j_{\textrm{eff}}=1/2$ and one $j_{\textrm{eff}}=3/2$ site, these terms allow the exciton to move.

Without the Jahn-Teller--mediated motion, i.e. for $H={H}^{\textrm{mag}}_{\rm SE}+
{H}^{\textrm{exc}}_{\rm SE}$, the problem was discussed in
Refs.~\onlinecite{PRL2012Kim, Naturecom2014Maria}. Exciton propagation
due to superexchange is analogous to the mechanism governing orbital excitations in
cuprates~\cite{Wohlfeld2011,Schlappa2012} and is strongly 
coupled to the magnon-like $j_{\textrm{eff}}=1/2$ excitations. We are going to show here that the Jahn-Teller coupling $\mathcal{H}_{\rm JT}(3/2, 1/2)$ provides an additional channel for delocalization
whose signatures can be clearly distinguished from the pure superexchange
scenario. 

\begin{figure}[!t]
   \subfigure[]{
     \includegraphics[width=0.95\linewidth]{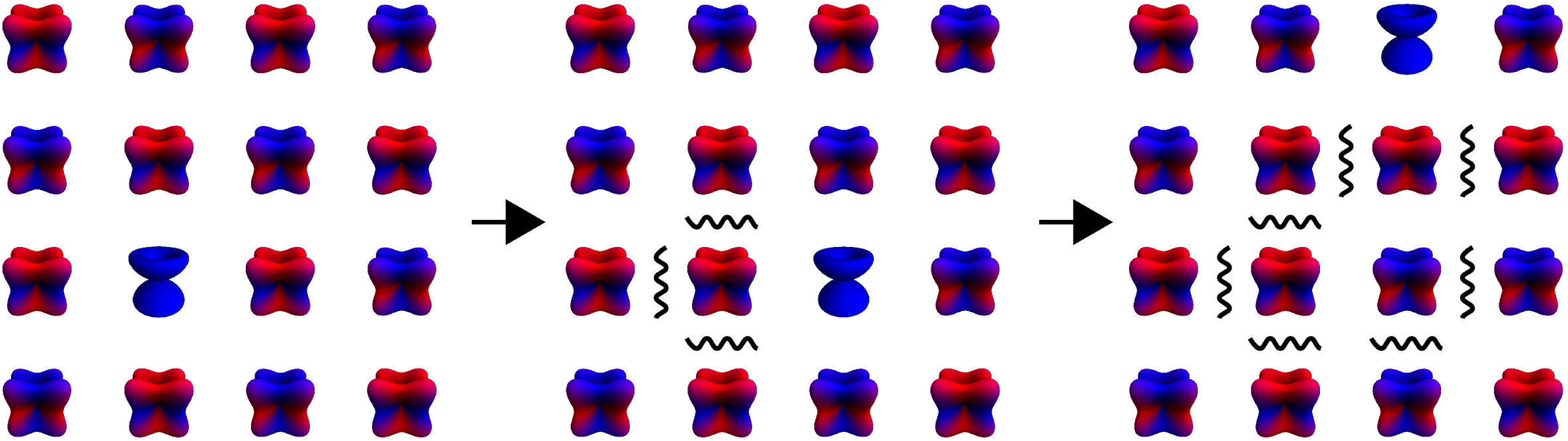}
   \label{cartoonHoppinga}}
 \subfigure[]{
 \includegraphics[width=0.95\linewidth]{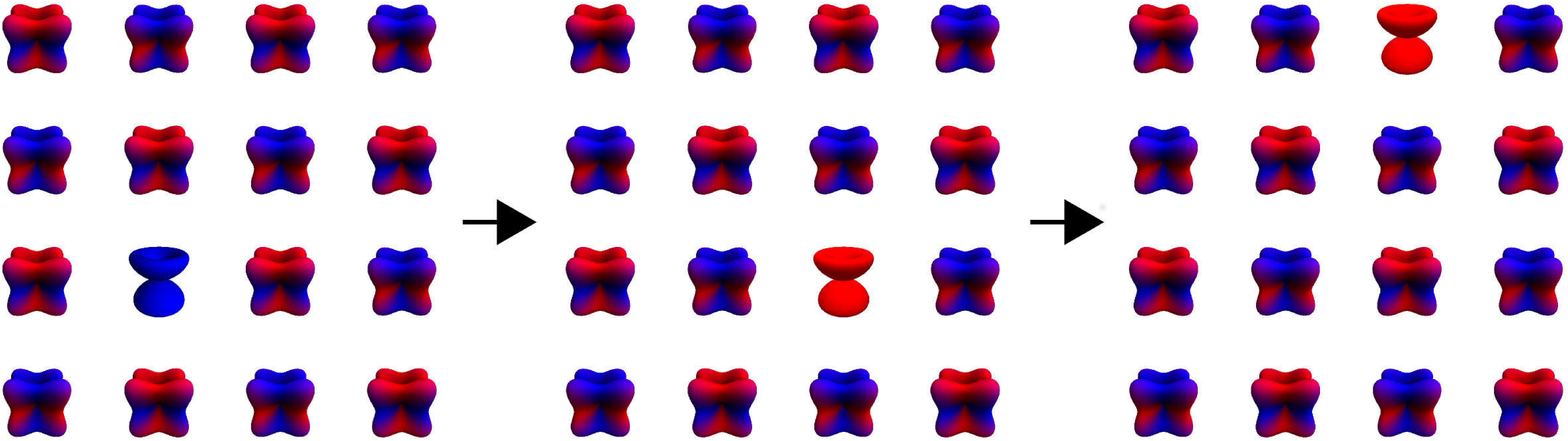}
    \label{cartoonHoppingb}}
      \caption{Cartoon showing the two types of nearest neighbor hopping of a $j_{\textrm{eff}}=3/2$ exciton in the antiferromagnetically--ordered background: 
(a) Polaronic hopping (due to Jahn-Teller effect $or$ superexchange): a $j_{\textrm{eff}}=3/2$ exciton with the $j_z=-3/2$ quantum number (left panel) does not change its $j_z$ quantum number
during the hopping process to the nearest neighbor sites (middle / right panels) and thus the $j_{\textrm{eff}}=1/2$ magnons are created at each step
of the excitonic hopping (wiggle lines on middle and right panels).
(b) Free hopping (solely due to Jahn-Teller effect): a $j_{\textrm{eff}}=3/2$ exciton with the $j_z=-3/2$ quantum number (left panel) hops to the nearest neighbor site 
and acquires $j_z=3/2$ quantum number (middle panel). Note that in this case the $j_{\textrm{eff}}=1/2$ magnons are {\it not} created in the system (middle / right panels).
\label{cartoonHopping}}
\end{figure}

Following Refs.~\cite{PRL2012Kim, Naturecom2014Maria}, we extend a scheme that was widely 
used to describe motion in an antiferromagnetic background~\cite{Martinez1991,Kane1989,Wohlfeld2009,Bala1995} in order
to include Jahn-Teller--mediated exciton motion. The scheme 
amounts to applying Holstein-Primakoff, Fourier and Bogoliubov transformations (see Ref.~\cite{SM} for details) to arrive
at the Hamiltonian
\begin{align}
	    \label{H_finalmag}
	{H}^{\textrm{mag}}_{\rm SE}&= \sum\limits_{{\bf k}}{\omega_{{\bf k}}a_{{\bf k}}^{\dag}a^{\phantom{\dagger}}_{{\bf k}}},\\
	{H}^{\textrm{exc}}_{\rm SE} + {H}^{\textrm{exc}}_{\rm JT}&=
        \sum\limits_{{\bf k}}{(  \hat{E}^{\rm SE}_{{\bf k}} +  \hat{E}^{\rm JT}_{{\bf k}} )
          \hat{\chi}^\dagger_{\bf k} \hat{\chi}^{\phantom{\dagger}}_{{\bf k}} }	\label{H_final2}\\
        + &\sum\limits_{{\bf k}, {\bf q}}{\left[(\hat{M}^{\rm SE}_{{\bf k},{\bf q}} + \hat{M}^{\rm JT}_{{\bf k},{\bf q}})
	\hat{\chi}^\dagger_{{\bf k}} \hat{\chi}^{\phantom{\dagger}}_{{\bf k}- {\bf q}}a^{\phantom{\dagger}}_{{\bf q}}+h.c.\right]}. \label{H_coupl} 	
\end{align}
Equation~(\ref{H_finalmag}) describes the isospin
`magnons' originating from ${H}^{\textrm{mag}}_{\rm SE}$, 
$a^\dag_{{\bf k}}$ creates a magnon with momentum ${\bf k}$ 
and energy $\omega_{{\bf k}}$, see Ref.~\cite{SM}. 
A free exciton hopping is included in Eq.~(\ref{H_final2}), it can
either be due to second- and third-neighbor
superexchange~\cite{Naturecom2014Maria}, or originate from coupling to
the lattice. Finally, Eq.~(\ref{H_coupl}) captures the
coupling between exciton hopping and the isospin background: Both
Jahn-Teller effect and superexchange can allow the exciton to exchange
place with a nearest-neighbor isospin without flipping said
isospin. This creates `faults' in the alternating order, see
Fig.~\ref{cartoonHoppinga}, and thus creates or annihilates magnons.  

Let us now discuss in more detail the contributions due to the
Jahn-Teller effect; for the pure superexchange problem, we refer to
Refs.~\cite{PRL2012Kim, Naturecom2014Maria}. 
The Jahn-Teller vertex $\hat{M}^{\rm JT}_{{\bf k}, {\bf q}}$ and the free excitonic 
dispersion $\hat{E}^{\rm JT}_{{\bf k}}$ are calculated here from $\mathcal{H}_{\rm JT}(3/2, 1/2)$
and read:
$ \hat{M}^{\rm JT}_{{\bf k}, {\bf q}}  =  z V \hat{m}^{\rm JT} \cdot | \gamma_{{\bf k}}v_{{\bf q}}+\gamma_{{\bf k}-{\bf q}}u_{{\bf q}} | / \sqrt{N}$
and $ \hat{E}^{\rm JT}_{{\bf k}}  = z V  \hat{e}^{\rm JT}  \cdot \left|\gamma_{\bf k} \right| $
where $N$ is the total number of sites, $z = 4$ is the coordination
number for a square lattice, and $\gamma_{{\bf k}}=\tfrac{1}{2}(\cos
k_x + \cos k_y)$. 
The Bogoliubov coefficients $u_{{\bf k}}, v_{{\bf k}}$ and the diagonal (off-diagonal) matrix $\hat{m}^{\rm JT}$ ($\hat{e}^{\rm JT}$) are explicitly given in Ref.~\cite{SM}.
The crucial new feature will turn out to come from the free dispersion
$\hat{E}^{\rm JT}_{{\bf k}}$, where the Jahn-Teller effect induces a
nearest-neighbor contribution absent from
superexchange. 

\begin{figure}[!t]
 \centering
\subfigure{
\includegraphics[width=0.467\linewidth]{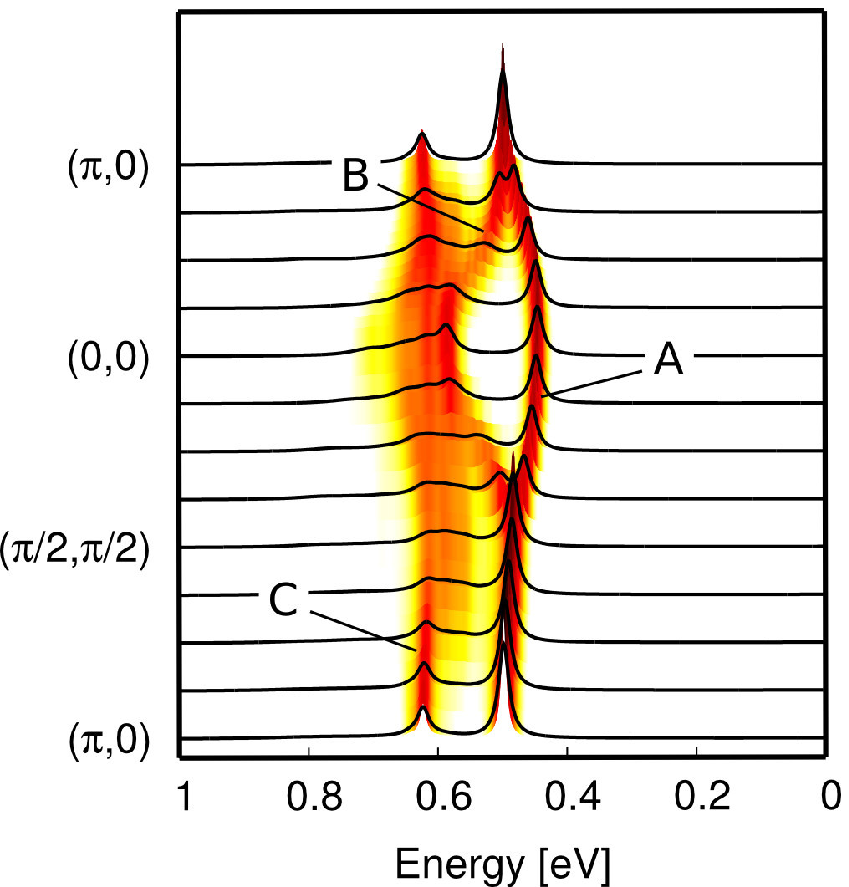}
\llap{
  \parbox[b]{2.3in}{\scriptsize{(a)}\\\rule{0ex}{1.5in}
  }}%
}
\subfigure{
\includegraphics[width=0.467\linewidth]{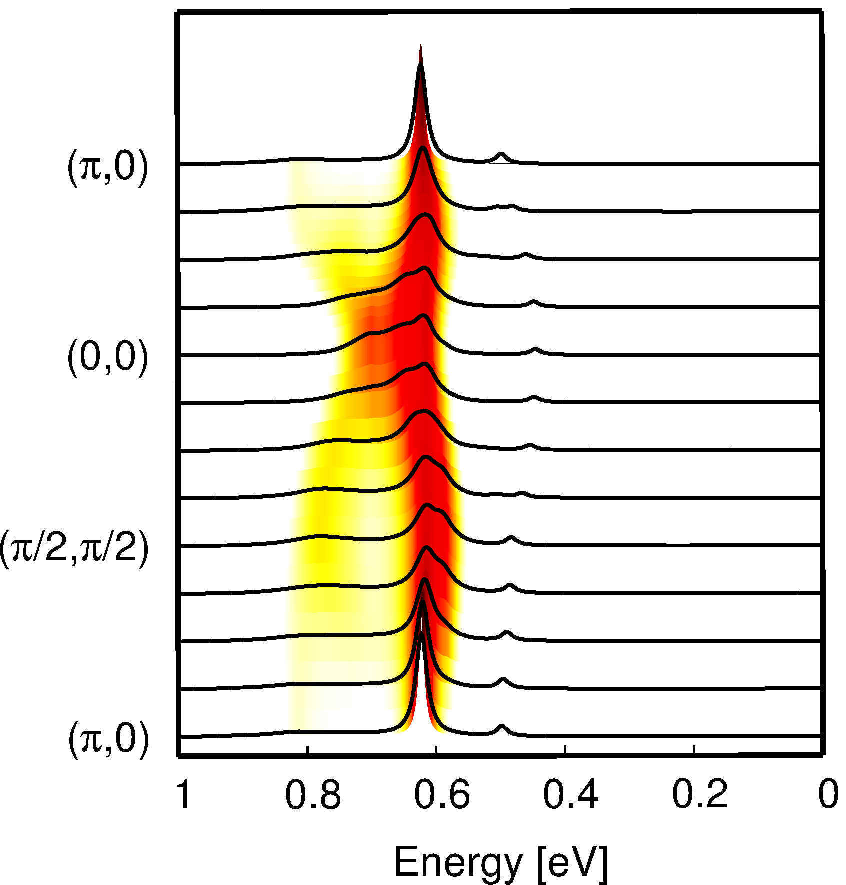}
\llap{
  \parbox[b]{2.3in}{\scriptsize{(b)}\\\rule{0ex}{1.5in}
  }}%
}
\caption{
\kw{Spin-orbit exciton with both superexchange and
  Jahn-Teller interaction, see (\ref{H_finalmag}) - (\ref{H_coupl})
  calculated using the SCBA. Intensities are given for two RIXS geometries:  (a) normal and (b) grazing incidence~\cite{Naturecom2014Maria}.
`A', `B', `C' in panel (a) denote three main features of the spectrum.
Jahn-Teller interaction $V = 0.8 J_1$ and broadening $\delta = 0.05 J_1$. 
Superexchange parameters $J_2 = -0.33 J_1 $, $J_3 = 0.25 J_1 $, $W_1= 0.5J_1$~\cite{Naturecom2014Maria}, 
and $W_2=W_3=0$. Following Ref.~\cite{Naturecom2014Maria} 
on-site energy of the exciton is $10J_1\approx \tfrac{3}{2}\lambda$, crystal-field splitting between $|j_z|=1/2$ and $|j_z| = 3/2$ states
is  $2.29 J_1$, and $J_1=0.06$ eV.}
\label{fig:main}} 
\end{figure}

\kw{{\it Results}}
We evaluate the Green function Eq.~(\ref{green}) using the self-consistent Born approximation (SCBA)
-- a diagrammatic approach that takes into account diagrams of
rainbow-type (see e.g. Ref.~\cite{Martinez1991}). 
\kw{The excitonic spectral functions
are calculated numerically for a $32\times 32$ cluster, taking into
account `matrix elements' depending on the angle of the incident beam~\cite{Naturecom2014Maria},  
 and shown in
Fig.~\ref{fig:main}. The most striking difference to the pure
superexchange scenario becomes visible in the so-called `normal' RIXS
geometry [cf. Fig.~\ref{fig:main}(a)]: a dispersive feature at around $0.4$ eV
(denoted as A in the figure) that has its minimal energy at ${\bf k} =
(0,0)$ and disperses upward towards the zone boundary, where it merges
with the B feature.  }

\md{An unexplained feature
with minimum at the $\Gamma$ point was observed in
normal-incidence RIXS experiments on Sr$_2$IrO$_4$
~\cite{Naturecom2014Maria}, albeit with a weaker intensity. This 
discrepancy may be due to (i) contributions to the RIXS intensity of the exciton beyond the one determined in the fast core-hole approximation ~\cite{Ament2011,Wohlfeld2015} or
(ii) the SCBA over-emphasizing the quasiparticle spectral
weight~\cite{Naturecom2014Maria}. Some fine-tuning of the unknown 
constant $V$ is needed to reproduce the experimental
dispersion, especially the merging with the B feature, see
Ref.~\cite{SM} for details. It is here worth noting that a similar peak was also seen in Na$_2$IrO$_3$~\cite{PhysRevLett.110.076402},
where it does not merge with the higher-energy features,
suggesting that the merging may be a detail specific to
Sr$_2$IrO$_4$. In contrast and as discussed below, the minimum at the
$\Gamma$ point is a robust and characteristic feature of 
Jahn-Teller--mediated propagation, because superexchange--driven peaks
invariably have a \emph{maximum} at the $\Gamma$ point. }

\begin{figure}[!t]
 \centering
\subfigure{
\includegraphics[width=0.467\linewidth]{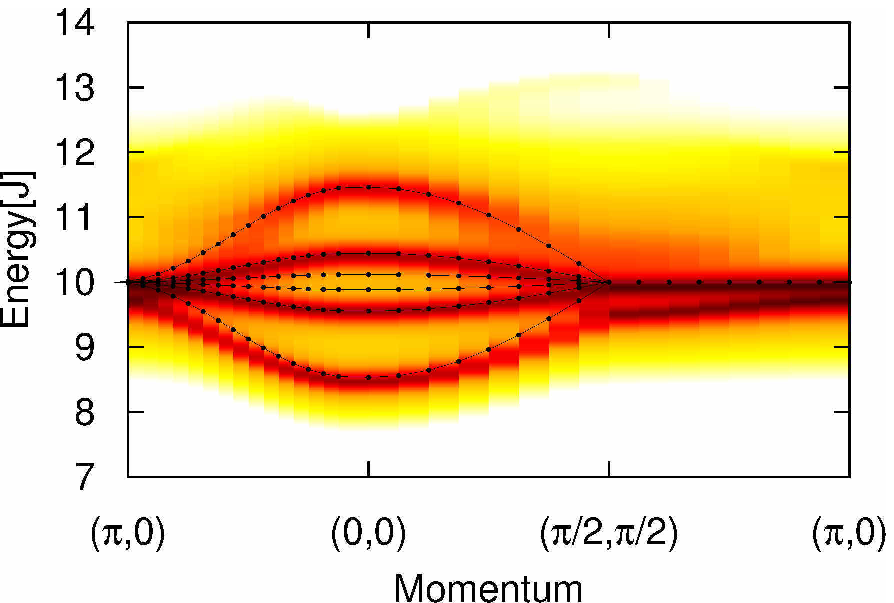}\llap{
  \parbox[b]{1.89in}{\scriptsize{(a) Jahn-Teller}\\\rule{0ex}{0.9in}
  }}\label{graphJT}
}
\subfigure{
\includegraphics[width=0.467\linewidth]{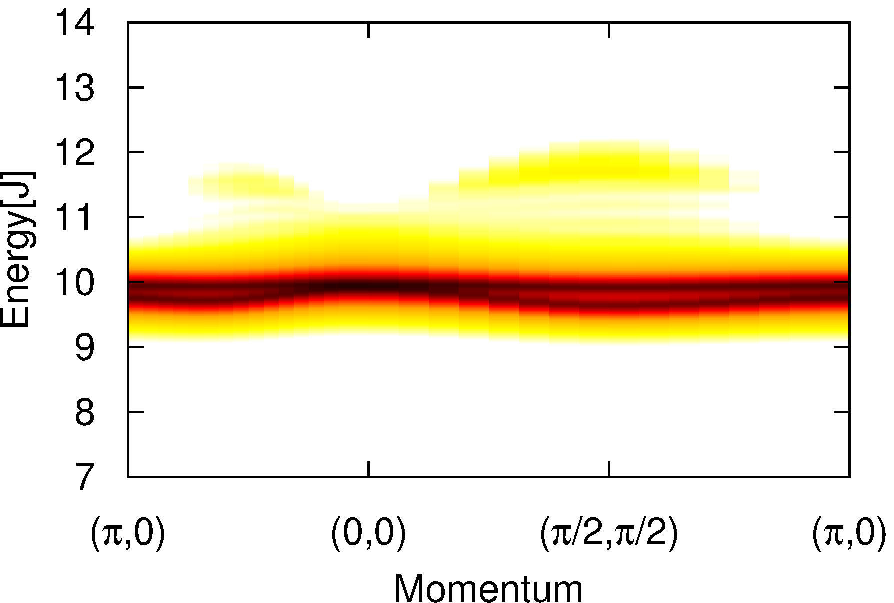}\llap{
  \parbox[b]{1.81in}{\scriptsize{(b) Jahn-Teller, \\$\hat{E}^{\rm JT}_{{\bf k}}\equiv 0$}\\\rule{0ex}{0.77in}
  }}\label{graphJTfree}
}
\subfigure{
\includegraphics[width=0.467\linewidth]{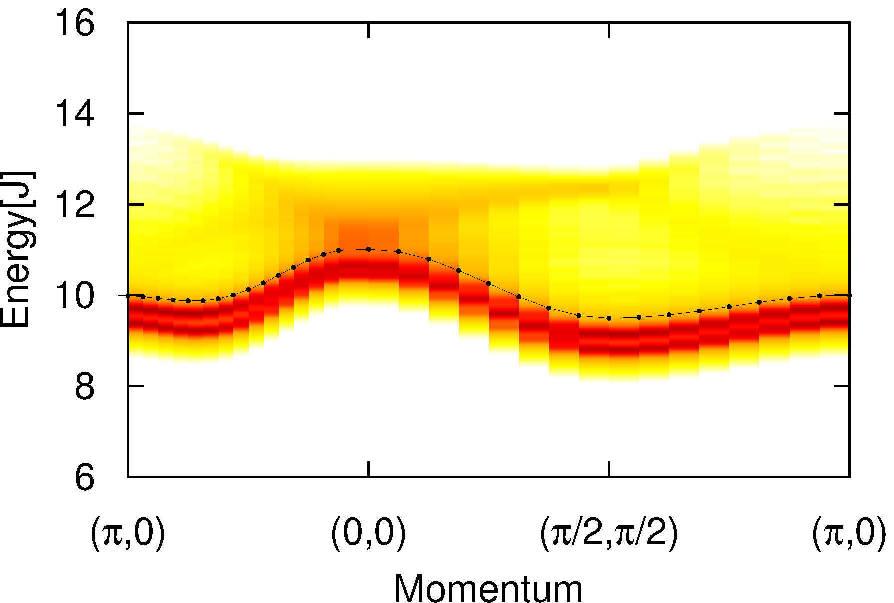}\llap{
  \parbox[b]{1.73in}{\scriptsize{(c) Superexchange}\\\rule{0ex}{0.9in}
  }}\label{graphSE}
  }
\subfigure{
\includegraphics[width=0.467\linewidth]{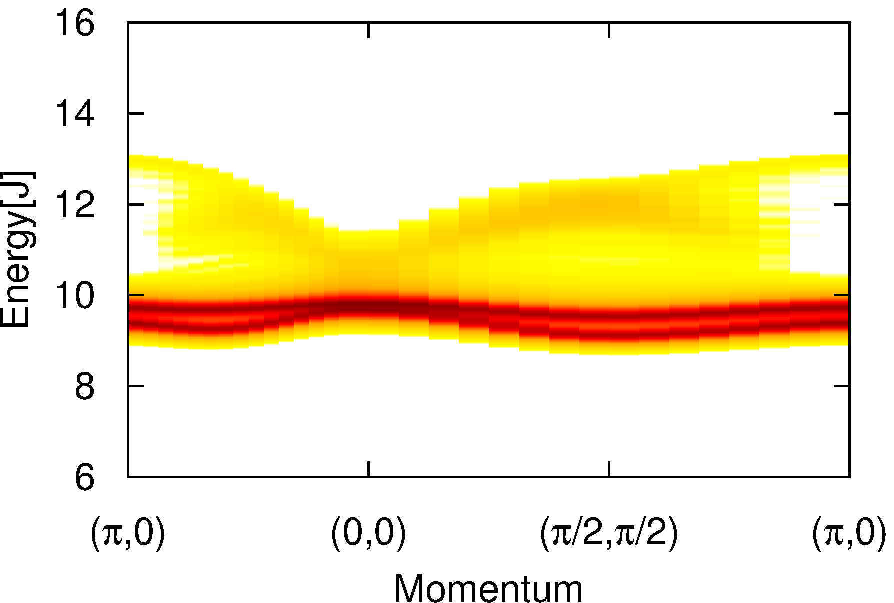}\llap{
  \parbox[b]{1.66in}{\scriptsize{(d) Superexchange, \\$\hat{E}^{\rm SE}_{{\bf k}}\equiv 0$}\\\rule{0ex}{0.77in}
  }}\label{graphSEfree}
}
\caption{
\md{Spin-orbit exciton spectra with propagation driven by either superexchange \emph{or} Jahn-Teller interaction [calculated using SCBA, see text]:
(a) Jahn-Teller only,
(b) Jahn-Teller only \emph{and} setting $\hat{E}^{\rm JT}_{{\bf k}} \equiv 0$,
(c) superexchange only,
(d) superexchange only \emph{and} setting $\hat{E}^{\rm SE}_{{\bf k}}
\equiv 0$.
For clarity, parameters are chosen slightly different from those used
in Fig.~\ref{fig:main}:  $J_2 = - 0.33 J_1 $, $J_3 = 0.25 J_1 $,
$W_1=0.5J_1$, $W_2 = W_3 = 0.13 J_1$~\cite{Naturecom2014Maria}, Jahn-Teller
interaction $V = J_1$ and broadening $\delta = 0.05 J_1$.  
Spectra are offset by the exciton energy of
$10J_1 \approx \tfrac{3}{2}\lambda$.
Dotted lines in (a) and (c) follow the free dispersion relations given by $\hat{E}^{\rm JT}_{{\bf k}}$ and $\hat{E}^{\rm SE}_{{\bf k}}$, respectively.}
\label{fig:comparison}} 
\end{figure}

\kw{{\it Discussion}}
Figure~\ref{fig:comparison} illustrates the qualitative difference
between Jahn-Teller and superexchange mediated exciton
propagation, with panels (a) and (c) showing the purely Jahn-Teller
(${H}^{\textrm{exc}}_{\rm SE} \equiv 0$) and purely
superexchange (${H}^{\textrm{exc}}_{\rm JT} \equiv 0$)
scenarios. A striking difference is that the two quasi-particle--like branches of the
superexchange case (c) become four in the Jahn-Teller case (a)  -- one of which has indeed a minimum at the
$\Gamma$ point.
We continue the
analysis by noting that both mechanisms
allow in principle for a `free' dispersion without disturbing the
alternating isospin order, see Eq.~(\ref{H_final2}), as
well as for a `polaronic' propagation involving magnons, see
Eq.~(\ref{H_coupl}). Panels (b) and (d) include only the latter 
and reveal that the two mechanisms are then almost
indistinguishable. This points to a dominant role for isospin
fluctuations (on the scale of $J$ in both scenarios) in the
`polaronic' part of exciton motion. 

This brings us to the following question: is the difference between
the free dispersion relation in the superexchange and in the
Jahn-Teller generic or it is just a matter of fine-tuning of the
parameters? It turns out that the difference between these two
dispersion relations is of fundamental nature. The crucial aspect
concerns the nearest-neighbor process, which is therefore depicted for
superexchange and Jahn-Teller effect in Fig.~\ref{cartoonHopping}. 
In superexchange, the exciton propagates by exchanging place with an
isospin while both conserve their `spin', i.e. their $j_z$ quantum number. In an alternating isospin order, where nearest
neighbors are always of opposite $j_z$, this necessarily
creates or removes `defects', see Fig.~\ref{cartoonHopping}(a), and
thus magnons. The Jahn-Teller effect, in contrast, allows the exciton
and the isospin to flip their quantum numbers while exchanging places and this allows for the nearest
neighbor hopping of an exciton without creating magnons, i.e.,  a free excitonic
dispersion. The origin of the difference is that the hole hopping
driving superexchange conserves the $j_z$ quantum number, while the
lattice-mediated Jahn-Teller effect is insensitive to the orbital
phase. This allows $j_z$ to change during Jahn-Teller--driven
propagation and accordingly yields four quasi-particles rather than two.

{\it Conclusions}
We analyzed here the impact of a lattice-mediated Jahn-Teller effect
in the presence of strong SOC, which 
quenches orbital degeneracy in the ground state. We found that 
the Jahn-Teller effect remains present for excited states, 
and in particular allows for  a `free'
nearest-neighbor hopping of the spin-orbit exciton without producing
defects in the alternating $j_{\textrm{eff}}=1/2$ ordering of the ground state. The
tell-tale spectral signature is a dispersion 
with a minimum at the $\Gamma$ point, which was observed in
experiment but cannot be explained with superexchange
alone~\cite{Naturecom2014Maria}. 
\md{Experiments  on Sr$_2$IrO$_4$  at
higher temperatures moreover reveal
an active orbital degree of freedom and its coupling to the lattice~\cite{Gretarsson2015}, corroborating the relevance
of  Jahn-Teller physics when going beyond the ground state.}

\md{We have found spin-orbit coupling to substantially affect the interplay of Jahn-Teller effect and
  superexchange. In $3d$ compounds with weak spin-orbit coupling and
unquenched orbital degeneracy (e.g. in manganites~\cite{PhysRevB.57.R5583,PhysRevB.59.3295}) both act on the same microscopic degree of freedom
(i.e. orbitals) and in general lead to similar signatures. In the
strongly spin-orbit--coupled $5d$ case, however, Jahn-Teller effect (determined purely by the orbital) and superexchange (strongly affected by spin-orbit
entanglement) address different microscopic degrees of freedom. Their
interplay is thus far more intricate, as is coupling between ions with and without strong spin-orbit
coupling~\cite{Brzezicki2015}.}

{\it Acknowledgments}
We are grateful to A.~M.~Ole\'s and N.~Bogdanov for fruitful
discussions and in particular wish to thank B. J. Kim for discussion
and helpful comments. The authors would all like to thank the staff of the Kavli Institute for Theoretical Physics (UCSB)
for their kind hospitality. This research was supported in part by the National Science Foundation under Grant No. NSF PHY11-25915
and the Deutsche Forschungsgemeinschaft (SFB 1143 and Emmy-Noether program) .
K.W. acknowledges support from the DOE-BES Division of Materials Sciences and Engineering (DMSE) under Contract No. DE-AC02-76SF00515 (Stanford/SIMES) and from the Polish National Science Center (NCN) under Project No. 2012/04/A/ST3/00331.  
\section{Supplemental Materials}
\subsection{A: Derivation of the polaronic Hamiltonian \\ from the Jahn-Teller model}
As discussed in the main text of the paper, the interaction between the orbital angular momenta as induced by the Jahn-Teller effect is described by the following 
Hamiltonian~\cite{Kugel1984}:
\begin{align}
\label{SMHJT}
\mathcal{H}_{\rm JT}&=V \sum\limits_{\langle {\bf i}, {\bf j} \rangle}\left[\left(l_{\bf i}^z\right)^2-\frac{2}{3}\right]\left[\left(l_{\bf j}^z\right)^2-\frac{2}{3}\right]  \nonumber \\
&+V \sum\limits_{\langle {\bf i},{\bf j} \rangle}  
\left[\left(l_{\bf i}^x\right)^2-\left(l_{\bf i}^y\right)^2\right]  
\left[\left(l_{\bf j}^x\right)^2-\left(l_{\bf j}^y\right)^2\right] 
\nonumber \\
&+ \kappa V \sum \limits_{\langle {\bf i},{\bf j} \rangle}\left[\left(l_{\bf i}^xl_{\bf i}^y+l_{\bf i}^yl_{\bf i}^x\right)(l_{\bf j}^xl_{\bf j}^y+l_{\bf j}^yl_{\bf j}^x)+...\right].
\end{align}
Here $V$ describes the Jahn-Teller interaction due to the coupling to the tetragonal modes, while
$\kappa V$ stands for the coupling between the trigonal modes. ${\bf l}$ is the orbital angular moment operator for 
the $t_{2_g}$ electrons (see also main text of the paper). 

In this part of the Supplemental Materials we show how to derive the polaronic Hamiltonian for the $j_{\textrm{eff}}=3/2$ excitons from the above Jahn-Teller
interaction -- we perform this derivation in two steps:

{\it Firstly}, since we are interested here in the effective interaction between the ${\bf j}={\bf l}+ {\bf s}$ spin-orbital angular momenta (as induced by the Jahn-Teller effect), 
we rewrite the above Jahn-Teller Hamiltonian in the basis spanned by eigenvectors of ${\bf j}^2$ and $j_z$ (the `${\bf j}$--basis'). 
Thus, we make a basis transformation from
the `${\bf l} \cdot {\bf s}$--basis' (with the effective $l_{\textrm{eff}}=1$ and $s=1/2$):
\begin{align}
\hat{A}=\left(|yz,\uparrow \rangle, |yz,\downarrow \rangle, |xz,\uparrow \rangle, |xz,\downarrow \rangle,
|xy,\uparrow \rangle, |xy,\downarrow \rangle \right),
\end{align}
in which the above Jahn-Teller Hamiltonian is written
into the `${\bf j}$--basis' (with the effective $j_{\textrm{eff}} = 1/2$ or $j_{\textrm{eff}} = 3/2$ and appropriate $j_z$ quantum numbers):
\begin{align}
\hat{J}= \left(|\frac{1}{2},\frac{1}{2}\rangle, |\frac{1}{2},\frac{-1}{2}\rangle,|\frac{3}{2},\frac{3}{2}\rangle,
|\frac{3}{2},\frac{1}{2}\rangle, |\frac{3}{2},\frac{-1}{2}\rangle, |\frac{3}{2},\frac{-3}{2}\rangle \right).
\end{align}
using the Clebsch-Gordon coefficients:\\
    \begin{equation}
    \label{clebshgordon}
      \hat{J}=\begin{pmatrix}
      0 & -\frac{1}{\sqrt{3}} & 0 & -\frac{i}{\sqrt{3}} & -\frac{1}{\sqrt{3}} & 0\\
      -\frac{1}{\sqrt{3}} & 0 & \frac{i}{\sqrt{3}} & 0 & 0 & \frac{1}{\sqrt{3}}\\
      -\frac{1}{\sqrt{2}} & 0 & -\frac{i}{\sqrt{2}} & 0 & 0 & 0\\
      0 & -\frac{1}{\sqrt{6}} & 0 & -\frac{i}{\sqrt{6}} & \sqrt{\frac{2}{3}} & 0\\
      \frac{1}{\sqrt{6}} & 0 & -\frac{i}{\sqrt{6}} & 0 & 0 & \sqrt{\frac{2}{3}}\\
      0 & \frac{1}{\sqrt{2}} & 0 & -\frac{i}{\sqrt{2}} & 0 & 0\\
      \end{pmatrix} \hat{A}. 
    \end{equation}
As a result we obtain the Jahn-Teller Hamiltonian which {\it a priori} consists of three distinct terms
\begin{align}
\mathcal{H}_{\rm JT} = \mathcal{H}_{\rm JT}(1/2, 1/2) + \mathcal{H}_{\rm JT}(3/2, 1/2)+ \mathcal{H}_{\rm JT}(3/2, 3/2),
\end{align}
as discussed already in detail in the main text.
Since we are interested here in the dynamics of the $j_{\textrm{eff}}=3/2$ exciton in the $j_{\textrm{eff}}=1/2$ alternating orbital background (see main text), 
we present here the explicit form of only  the $\mathcal{H}_{\rm JT}(3/2,1/2)$ part of the Hamiltonian:
   \begin{equation}
	\label{HJT2}
	\mathcal{H}_{\rm JT}(3/2, 1/2) \equiv \mathcal{H}_{\rm JT}^{exc}=\mathcal{H}_1+\mathcal{H}_2+\mathcal{H}_3, 
    \end{equation}
where
      \begin{flalign}
	    \mathcal{H}_1&=\frac{2V}{9}\sum\limits_{\langle {\bf i}, {\bf j} \rangle}{\left(c_{ {\bf i} \uparrow}^\dag  a_{ {\bf i} \uparrow} a_{ {\bf j} \uparrow}^\dag c_{ {\bf j} \uparrow} +
	c_{ {\bf i} \downarrow}^\dag  a_{ {\bf i} \downarrow} a_{ {\bf j} \downarrow}^\dag c_{ {\bf j} \downarrow}+ h.c.\right)} \nonumber \\
	&-\frac{2V}{9}\sum\limits_{\langle {\bf i}, {\bf j} \rangle}{\left(c_{ {\bf i} \downarrow}^\dag  a_{ {\bf i} \downarrow} a_{ {\bf j} \uparrow}^\dag c_{ {\bf j} \uparrow} +
	c_{ {\bf i} \uparrow}^\dag a_{ {\bf i} \uparrow} a_{ {\bf j} \downarrow}^\dag  c_{ {\bf j} \downarrow}+ h.c.\right)} \nonumber\\
	&+\kappa V\sum\limits_{\langle {\bf i}, {\bf j} \rangle}{\left(c_{ {\bf i} \uparrow}^\dag a_{ {\bf i} \downarrow} a_{ {\bf j} \downarrow}^\dag c_{ {\bf j} \uparrow} +
	c_{ {\bf i} \downarrow}^\dag a_{ {\bf i} \uparrow} a_{ {\bf j} \uparrow}^\dag c_{ {\bf j} \downarrow}+ h.c.\right)},\\		
	    \mathcal{H}_2&=\frac{\kappa V}{3}\sum\limits_{\langle {\bf i}, {\bf j} \rangle}{\left(f_{ {\bf i} \uparrow}^\dag a_{ {\bf i} \uparrow} a_{ {\bf j} \uparrow}^\dag f_{ {\bf j} \uparrow} +
	f_{ {\bf i} \downarrow}^\dag  a_{ {\bf i} \downarrow} a_{ {\bf j} \downarrow}^\dag f_{ {\bf j} \downarrow}+ h.c.\right)} \nonumber \\
	&+\frac{2V(1+\kappa)}{3}\sum\limits_{\langle {\bf i}, {\bf j} \rangle}{\left(f_{ {\bf i} \downarrow}^\dag a_{ {\bf i} \uparrow} a_{ {\bf j} \uparrow}^\dag f_{ {\bf j} \downarrow}
	+f_{ {\bf i} \uparrow}^\dag a_{ {\bf i} \downarrow} a_{ {\bf j} \downarrow}^\dag f_{ {\bf j} \uparrow}+ h.c.\right)} \nonumber\\
	&+\frac{2V(\kappa-1)}{3}\sum\limits_{\langle {\bf i}, {\bf j} \rangle}{\left(f_{ {\bf i} \downarrow}^\dag a_{ {\bf i} \uparrow} a_{ {\bf j} \downarrow}^\dag f_{ {\bf j} \uparrow} +
	f_{ {\bf i} \uparrow}^\dag a_{ {\bf i} \downarrow} a_{ {\bf j} \uparrow}^\dag f_{ {\bf j} \downarrow}+ h.c.\right)},\\
	    \mathcal{H}_{3}&=-\frac{\kappa V}{\sqrt{3}}\sum\limits_{\langle {\bf i}, {\bf j} \rangle}{\left(
	f_{ {\bf i} \uparrow}^\dag a_{ {\bf i} \uparrow} a_{ {\bf j} \downarrow}^\dag c_{ {\bf j} \uparrow} +
	c_{ {\bf i} \uparrow}^\dag a_{ {\bf i} \downarrow} a_{ {\bf j} \uparrow}^\dag f_{ {\bf j} \uparrow}+h.c.\right)} \nonumber \\
	&-\frac{\kappa V}{\sqrt{3}}\sum\limits_{\langle {\bf i}, {\bf j} \rangle}{\left(
	c_{ {\bf i} \downarrow}^\dag a_{ {\bf i} \uparrow} a_{ {\bf j} \downarrow}^\dag f_{ {\bf j} \downarrow}
	+f_{ {\bf i} \downarrow}^\dag a_{ {\bf i} \downarrow} a_{ {\bf j} \uparrow}^\dag c_{ {\bf j} \downarrow} +h.c.\right)}. 
      \end{flalign}
Here $a^\dagger_{ {\bf i} \sigma}$ denotes an operator creating a hole on site ${\bf i}$ in the doublet carrying effective total momentum $j_{\textrm{eff}}=1/2$ and $\sigma \equiv j_z = \pm 1/2$,
while $c^\dagger_{ {\bf i} \sigma}$ ($f^\dagger_{ {\bf i} \sigma}$) are operators creating a hole on site ${\bf i}$ in the $j_{\textrm{eff}}=3/2$ quartet with $\sigma \equiv j_z= \pm 1/2, \pm 1/2$.

{\it Secondly}, we map the above Hamiltonian $\mathcal{H}_{\rm JT}^{exc}$ onto a polaronic model (see main text of the paper for the motivation).
We follow Ref.~\cite{Martinez1991} and perform the transformations:

(i) Since we assume that the ground state has antiferromagnetic order, we are allowed to rotate all isospins on one of the two antiferromagnetic sublattices:    
\begin{equation}
      \label{spinrotation}
      a_{ {\bf j} \sigma} \rightarrow a_{ {\bf j} -\sigma} \quad
            c_{ {\bf j} \sigma} \rightarrow c_{ {\bf j} -\sigma} \quad
                  f_{ {\bf j} \sigma} \rightarrow f_{ {\bf j} -\sigma}.
    \end{equation}
 
 (ii)  We introduce the magnon creation $\alpha^\dagger_{ {\bf i}}$ and spin-orbit exciton creation $\chi^\dagger_{ {\bf i},\alpha}$ operators (which are bosons and hard-core bosons, respectively).
We perform the Holstein-Primakoff transformation and substitute: 
     \begin{align}
	\label{HolPrimJT}
	c^\dagger_{ {\bf i} \uparrow}b_{ {\bf i} \uparrow}\rightarrow \chi^\dagger_{ {\bf i} B},  c^\dagger_{ {\bf i} \downarrow}b_{ {\bf i} \uparrow}& \rightarrow \chi^\dagger_{ {\bf i} C},\\
	c^\dagger_{ {\bf i} \uparrow}b_{ {\bf i} \downarrow}\rightarrow \chi^\dagger_{ {\bf i} B} \alpha_{\bf i}, c^\dagger_{ {\bf i} \downarrow}b_{ {\bf i} \downarrow}& \rightarrow \chi^\dagger_{ {\bf i} C}\alpha_{ {\bf i}},\\
	b^\dagger_{ {\bf i} \uparrow}c_{ {\bf i} \uparrow}\rightarrow \chi_{ {\bf i} B},  b^\dagger_{ {\bf i} \uparrow}c_{ {\bf i} \downarrow}& \rightarrow \chi_{ {\bf i} C},\\
	b^\dagger_{ {\bf i} \downarrow}c_{ {\bf i} \uparrow}\rightarrow \alpha_i^{\dag}\chi_{ {\bf i} B}, b^\dagger_{ {\bf i} \downarrow}c_{ {\bf i} \downarrow}& \rightarrow \alpha^{\dagger}_{ {\bf i}}\chi_{ {\bf i} C},
      \end{align}
and
      \begin{flalign}
	\label{HolPrimJTb}
	f^\dagger_{ {\bf i} \uparrow}b_{ {\bf i} \uparrow}\rightarrow \chi^\dagger_{ {\bf i} D},  f^\dagger_{ {\bf i} \downarrow}b_{ {\bf i} \uparrow}& \rightarrow \chi^\dagger_{ {\bf i} F},\\
	f^\dagger_{ {\bf i} \uparrow}b_{ {\bf i} \downarrow}\rightarrow \chi^\dagger_{ {\bf i} D}\alpha_{\bf i}, f^\dagger_{ {\bf i} \downarrow}b_{ {\bf i} \downarrow}& \rightarrow \chi^{\dagger}_{ {\bf i} F}\alpha_{ {\bf i}},\\
	b^\dagger_{ {\bf i} \uparrow}f_{ {\bf i} \uparrow}\rightarrow \chi_{ {\bf i} D},  b^{\dagger}_{ {\bf i} \uparrow}f_{ {\bf i} \downarrow}& \rightarrow \chi_{ {\bf i} F},\\
	b^\dagger_{ {\bf i} \downarrow}f_{ {\bf i} \uparrow}\rightarrow \alpha_i^{\dag}\chi_{ {\bf i} D}, b^\dagger_{ {\bf i} \downarrow}f_{ {\bf i} \downarrow}& \rightarrow \alpha^{\dagger}_{ {\bf i}}\chi_{ {\bf i} F}.
      \end{flalign}
Here B, C, D, F denote $j_z=1/2$,$j_z=-1/2$,$j_z=3/2$,$j_z=-3/2$ quantum numbers, respectively.

(iii) We perform the Fourier and Bogolyubov transformations (see e.g. Ref.~\cite{Martinez1991}): 
    \begin{equation}
      \begin{split}
      \label{Bogoliubov}
      a_{{\bf q}} = u_{{\bf q}} \alpha_{{\bf q}} -v_{{\bf q}} \alpha_{-{\bf q}}^\dag,\\
      a_{{\bf -q}}^\dag = u_{{\bf -q}} \alpha_{{\bf -q}}^\dag- v_{{\bf q}} \alpha_{{\bf q}}
      \end{split}
    \end{equation}
where the magnon energy $\omega_{{\bf q}}=\sqrt{A_{{\bf q}}^2-B_{{\bf q}}^2}$ and Bogolyubov coefficients
$u_{{\bf q}}$, $v_{{\bf q}}$ are given by the usual expressions in the linear spin-wave theory:
\begin{equation}
\begin{split}
      \label{Bogcoef3}
      u_{{\bf q}} = \frac{1}{\sqrt{2}}\sqrt{\frac{A_{{\bf q}}}{\omega_{{\bf q}}}+1},
      v_{{\bf q}} = -\frac{sign(B_{{\bf q}})}{\sqrt{2}}\sqrt{\frac{A_{{\bf q}}}{\omega_{{\bf q}}}-1},
\end{split}
\end{equation}
where the coefficients $A_{\bf q}$ and $B_{\bf q}$
are defined in a usual way, see e.g. Eq.~(8) in the Supplementary Material of 
Ref.~\cite{Naturecom2014Maria}. 
Here we neglected terms comprising two magnon operators, since it was shown that coupling to two magnons 
does not significantly change the polaronic spectrum (see for example  Ref.~\cite{Bala1995}).

After applying the above transformations to the Hamiltonian (\ref{HJT2}) we arrive at the following polaronic Hamiltonian for the propagation of the $j_{\textrm{eff}}=3/2$ spin-orbit exciton (see also main text of the paper):
\begin{align}
	\label{H_final}
	 {H}^{exc}_{\rm JT}=
\sum\limits_{{\bf k}, {\bf q}}{[\hat{M}^{\rm JT}_{{\bf k},{\bf q}}
	\hat{\chi}^\dagger_{{\bf k}} \hat{\chi}_{{\bf k}- {\bf q}}a_{{\bf q}}+h.c.]} + \sum\limits_{{\bf k}}{\hat{E}^{\rm JT}_{{\bf k}}\hat{\chi}^\dagger_{\bf k} \hat{\chi}_{{\bf k}} },
\end{align}
with the momentum-dependent vertices$  \hat{M}^{\rm JT}_{{\bf k}, {\bf q}}  =  z V \hat{m}^{\rm JT} \cdot | \gamma_{{\bf k}}v_{{\bf q}}+\gamma_{{\bf k}-{\bf q}}u_{{\bf q}} | / \sqrt{N}$
and $ \hat{E}^{\rm JT}_{{\bf k}}  = z V  \hat{e}^{\rm JT}  \cdot \left|\gamma_{\bf k} \right|$.
Here $\gamma_{{\bf q}}=\frac{1}{2}\sum\limits_{{\bf r}}{\cos {\bf q} \cdot {\bf r}}$ and the diagonal (off-diagonal) matrix $\hat{m}^{\rm JT}$ ($\hat{e}^{\rm JT}$)
describes the polaronic (free) hopping reads:
     \begin{align}
     \label{matrix}
      \hat{m}^{\rm JT}&= \frac{1}{3}\begin{pmatrix}
      \frac{1}{3}+\frac{\kappa}{2} &  0 & 0 & 0\\
      0 & \frac{1}{3}+\frac{\kappa}{2} & 0 & 0 \\
      0  & 0 & 1+\frac{\kappa}{2} &  0  \\
      0 & 0 & 0  & 1+\frac{\kappa}{2} \\
      \end{pmatrix},\\
      \end{align}
and
\begin{align}
     \label{matrix2}
     \hat{e}^{\rm JT}&=\frac{2}{3}
     \begin{pmatrix}
      0 &  -\frac{1}{3} & -\frac{\sqrt{3}\kappa}{2}  & 0\\
      -\frac{1}{3} &  0 & 0 &  -\frac{\sqrt{3}\kappa}{2}  \\
      -\frac{\sqrt{3}\kappa}{2}   & 0 & 0 &  -1 +\kappa \\
      0 & -\frac{\sqrt{3}\kappa}{2}  &  -1 +\kappa  & 0 \\
      \end{pmatrix},
    \end{align}
which acts on the row of
kets of the 'excited states' $X  = \left( |j_z = 1/2\rangle,  |j_z = -1/2\rangle, 
|j_z = 3/2\rangle, |j_z = -3/2\rangle \right)$ with $j_{\textrm{eff}}=3/2$.

\kw{\subsection{B: dependence of the results on the model parameters}}

\begin{figure}[!t]
 \centering
\subfigure{
\includegraphics[width=0.465\linewidth]{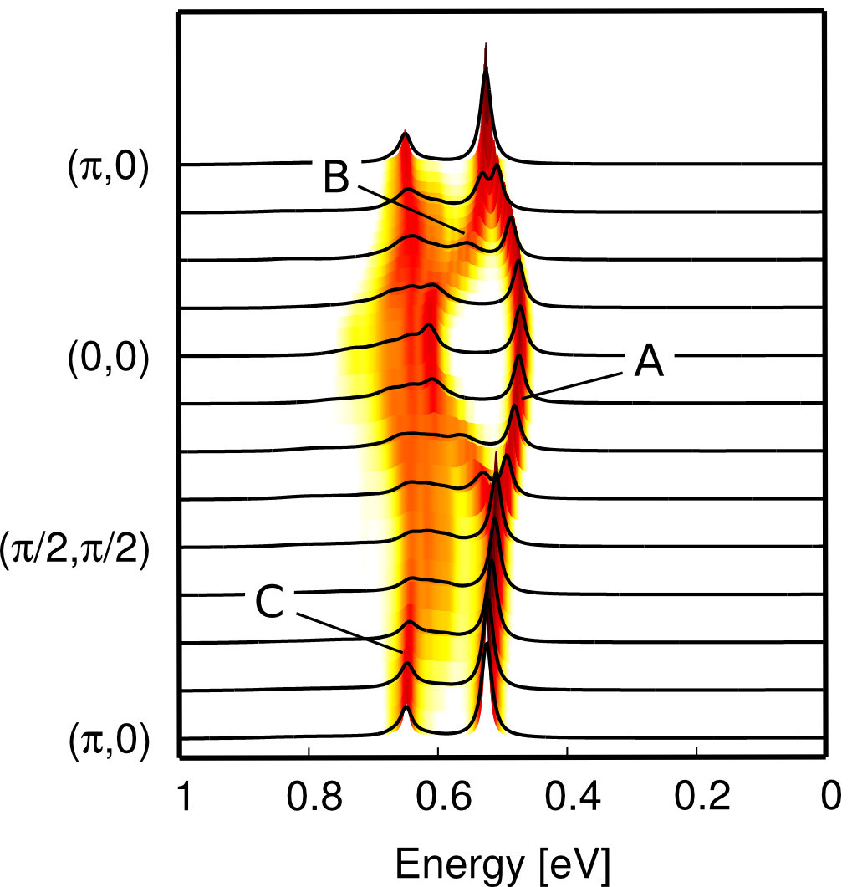}\llap{
  \parbox[b]{2.75in}{\scriptsize{(a) }\\\rule{0ex}{1.5in}
  }}\label{SMfig:comparison:a}
}
\subfigure{
\includegraphics[width=0.465\linewidth]{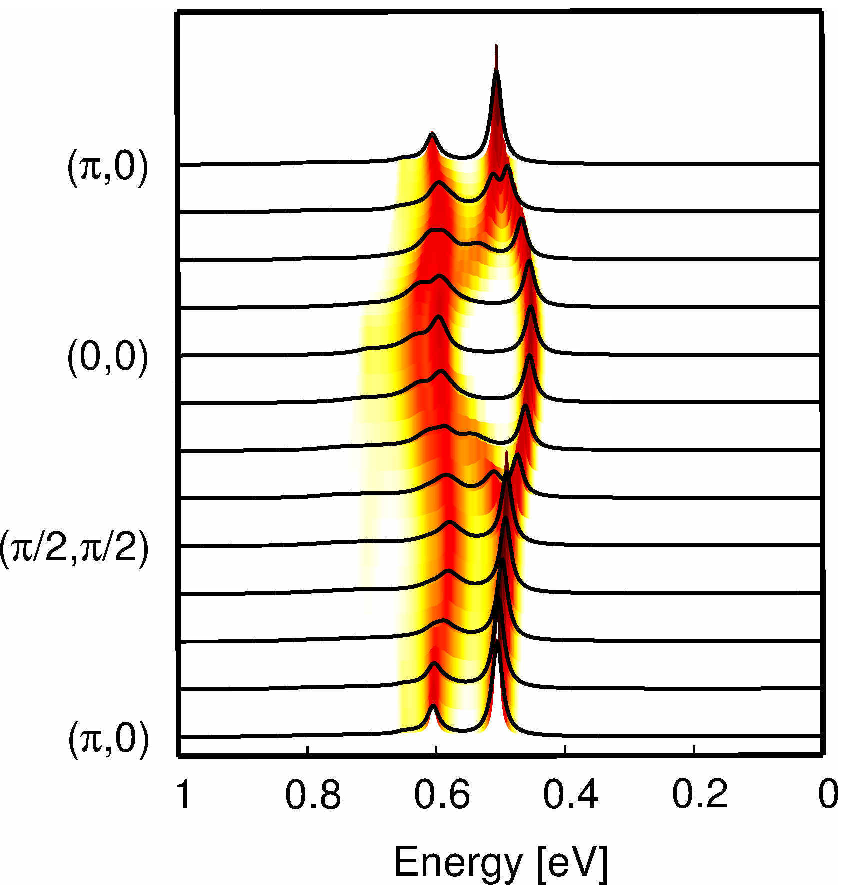}\llap{
  \parbox[b]{2.75in}{\scriptsize{(b)}\\\rule{0ex}{1.48in}
  }}\label{SMfig:comparison:b}
}
\subfigure{
\includegraphics[width=0.465\linewidth]{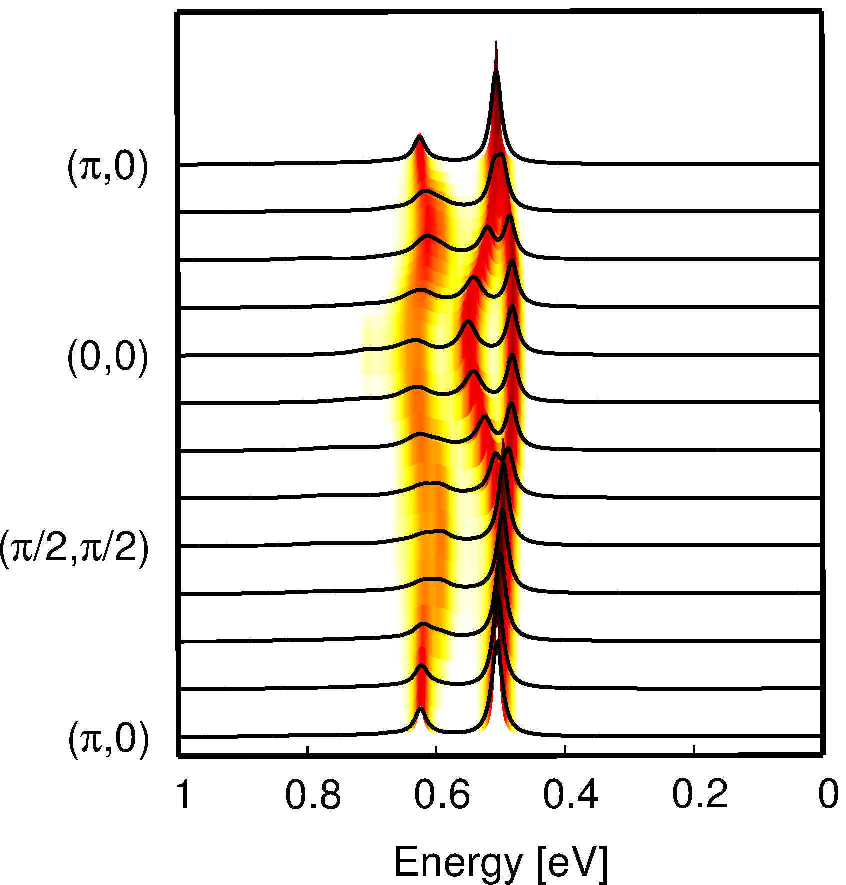}\llap{
  \parbox[b]{2.75in}{\scriptsize{(c)}\\\rule{0ex}{1.5in}
  }}\label{SMfig:comparison:c}
  }
\subfigure{
\includegraphics[width=0.465\linewidth]{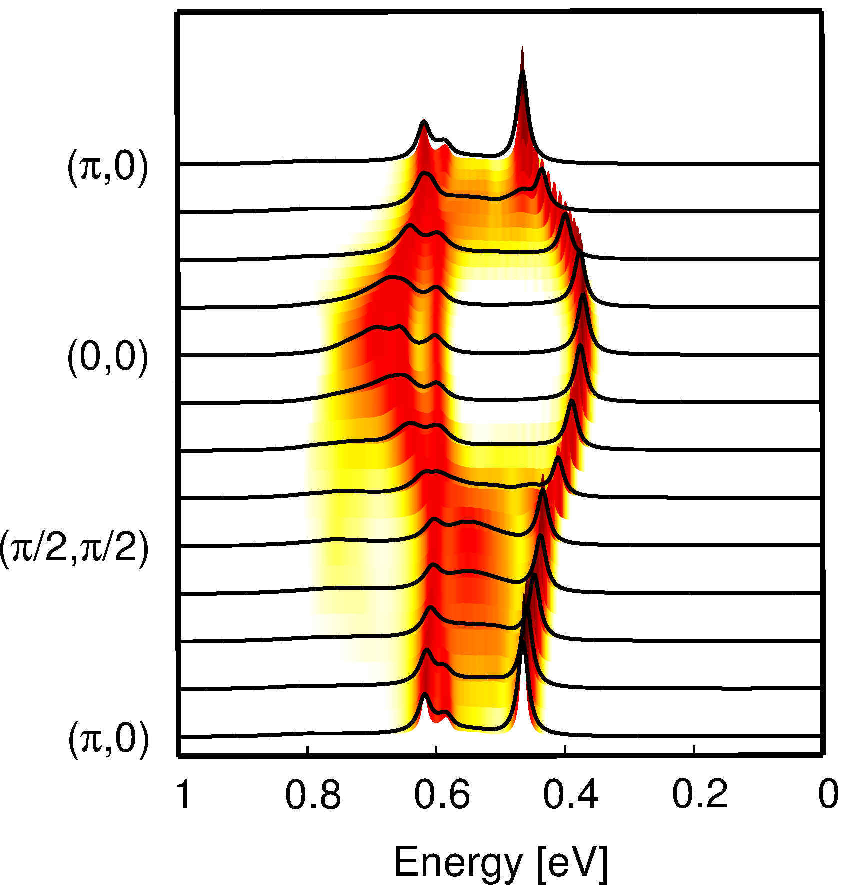}\llap{
  \parbox[b]{2.75in}{\scriptsize{(d)}\\\rule{0ex}{1.48in}
  }}\label{SMfig:comparison:d}
}
\subfigure{
\includegraphics[width=0.465\linewidth]{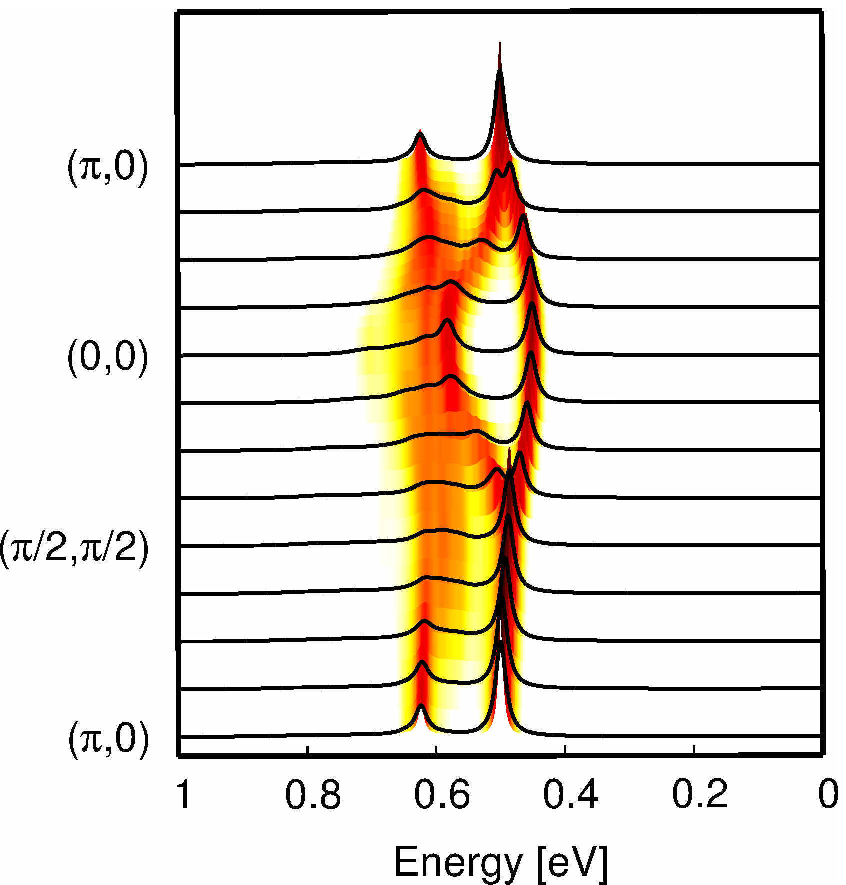}\llap{
  \parbox[b]{2.75in}{\scriptsize{(e)}\\\rule{0ex}{1.5in}
  }}\label{SMfig:comparison:e}
}
\subfigure{
\includegraphics[width=0.465\linewidth]{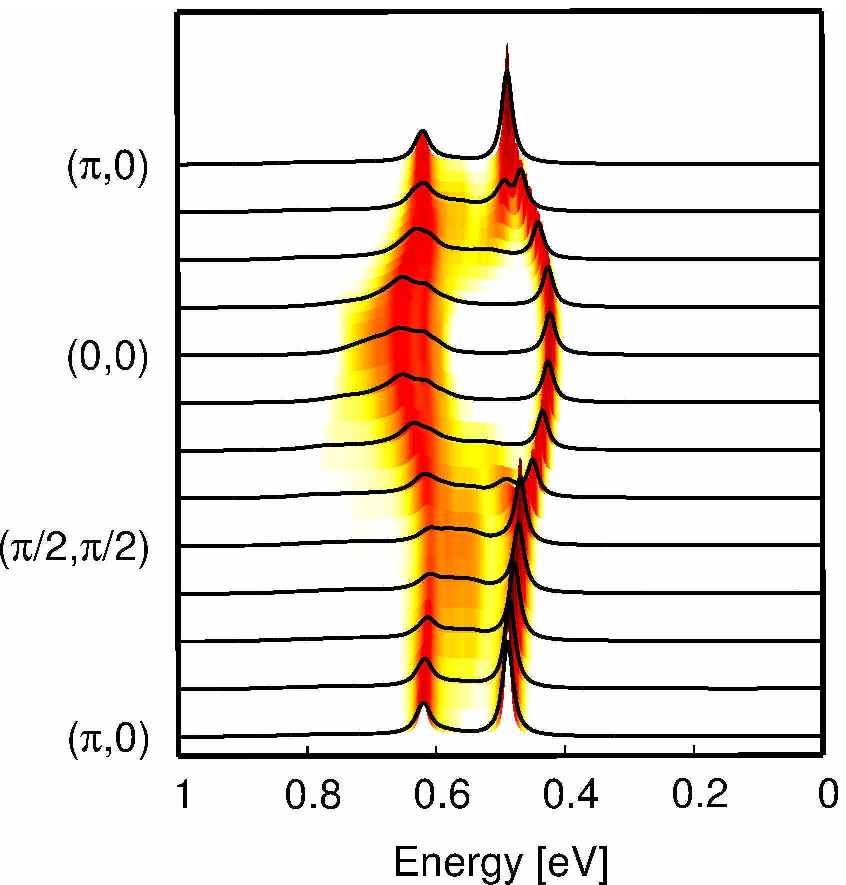}\llap{
  \parbox[b]{2.75in}{\scriptsize{(f)}\\\rule{0ex}{1.48in}
  }}\label{SMfig:comparison:f}
  }
\caption{\kw{Dependence of the spin-orbit exciton spectral function on the parameters of the model [Eq. (4) in the main text of the paper]: 
on-site spin-orbit coupling $\lambda$,
the on-site energy gap between the $|j_z|=1/2$ and $|j_z| = 3/2$ excitons $\Delta_{BC}$, and the Jahn-Teller coupling constant $V$ and $\kappa$:
(a) $\lambda=6.67 J_1$,
(b) $\Delta_{BC}= 1.86 J_1 $,
(c) $V = 0.4 J_1$,
(d) $V = 1.6 J_1$,
(e) $\kappa=0.025$,
(f) $\kappa=0.4$.
If not specified above, all other model parameters are as in the main text of the paper (cf. caption of Fig. 2).
Letters `A', `B', `C' in panel (a) denote three main spectral features of the spectrum -- see text for further details.} 
\label{SMfig:comparison}} 
\end{figure}%
\kw{In this part of the supplemental materials we show how the spectral function of the spin-orbit exciton calculated within our model depends on the model parameters: the on-site spin-orbit coupling $\lambda$, the on-site energy gap between the $|j_z|=1/2$ and $|j_z| = 3/2$ excitons (following the notation used in Ref.~\cite{Naturecom2014Maria} we call it $\Delta_{BC}$ below), and the Jahn-Teller coupling constants $V$ and $\kappa$. [The results for different choices of the superexchange parameters can already be inferred from Refs.~\cite{PRL2012Kim, Naturecom2014Maria}.]}

\kw{In Fig.~\ref{SMfig:comparison:a} the excitonic spectrum is shown for the value of $\lambda=6.67 J_1$ (which corresponds to one of the proposed values of $\lambda=400$ meV~\cite{Kim2008} for Sr$_2$IrO$_4$). We see that increasing the value of the spin-orbit coupling with respect to the one chosen in the main text of the paper leads to a merely modest shift of the spectral weight to higher energies without a significant change of the shape of the spectra. The decrease of  the on-site energy gap between the $|j_z|=1/2$ and $|j_z| = 3/2$ excitons from its main-text value of $\Delta_{BC} = 2.29 J_1$ to $\Delta_{BC} = 1.86 J_1$ (which follows from the crystal field splitting $\Delta = -155 $ meV as suggested for Sr$_2$IrO$_4$  by e.g. Ref.~\cite{Bogdanov2015}), cf. Fig.~\ref{SMfig:comparison:b}, 
leads to a small shift of the spectrum and also slightly renormalises the spectral weight, especially around $(\pi, 0)$.} 

\kw{In Fig.~\ref{SMfig:comparison:c}-\ref{SMfig:comparison:f} the dependence of the excitonic spectrum on the Jahn-Teller coupling constants is shown.
Since the values of the Jahn-Teller coupling constants are rather hard to estimate and to the best of our knowledge no estimates are available for Sr$_2$IrO$_4$, 
we vary these values in a rather wide range.
First we take $V$ twice smaller than the one used in the main text of the paper, $V=0.4 J_1$, and keep $\kappa$ unchanged. As we see in Fig.~\ref{SMfig:comparison:c}
such change affects the spectra in the following way: the middle feature [denoted as `B' in Fig.~\ref{SMfig:comparison:a}] shifts to the lower energies,
separates more from the the highest one [denoted as `C' in Fig.~\ref{SMfig:comparison:a}], and forms a clear maximum at the $\Gamma$ point.
Next, if we make $V$ twice larger w.r.t. the value suggested in the main text of the paper [see Fig.~\ref{SMfig:comparison:d}], then the effect is exactly opposite: feature B shifts to higher energies, almost merges with C and some spectral weight shifts from feature B to C. 
Finally, as one varies $\kappa$, one sees almost no changes for a smaller value of
$\kappa$ w.r.t. the value suggested in the main text of the paper [see Fig.~\ref{SMfig:comparison:e}], while for a relatively large $\kappa$ there is a relatively large shift of the spectral weight from feature B to C. It should also be noted that increasing the strength of the Jahn-Teller couplings (by making either $V$ {\it or} $\kappa$ larger) leads to a larger dispersion relation of all the features.} 

\kw{Altogether we conclude that there are rather severe constraints on the possible realistic values of these parameters, provided that the spectrum is intended to describe the excitonic propagation in one of the quasi-2D iridates (such as e.g. Sr$_2$IrO$_4$).
Moreover, the changes in the excitonic spectrum, due to the small variations in $\lambda$ or $\Delta_{BC}$, are rather small. On the other hand, the values of the Jahn-Teller constants in the iridium oxides are rather hard to estimate and the large variations in the values of the Jahn-Teller constants may indeed lead to some more substantial changes in the shape of the excitonic spectrum. Nevertheless, such changes are never as substantial as to completely alter the main qualitative features of the excitonic spectrum: the mere existence of the three main features (A, B, C) as well as the generic features of their dispersion relations.} 

\subsection{C: Understanding the free excitonic hopping arising from the Jahn-Teller model}

\begin{figure}
 \centering
\subfigure{
\includegraphics[width=0.465\linewidth]{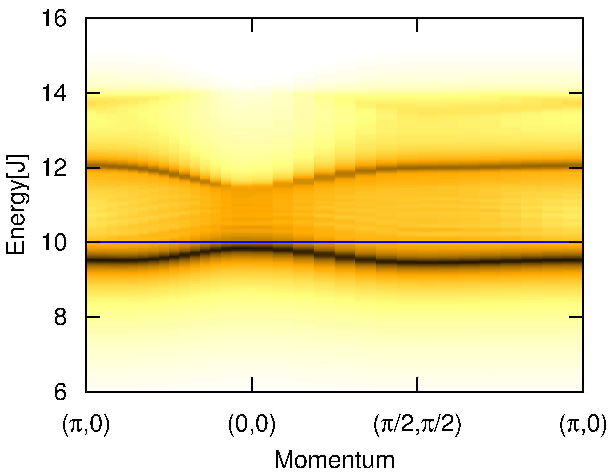}\llap{
  \parbox[b]{2.45in}{(a)\\\rule{0ex}{1.0in}
  }}\label{2a}
}
\subfigure{
\includegraphics[width=0.465\linewidth]{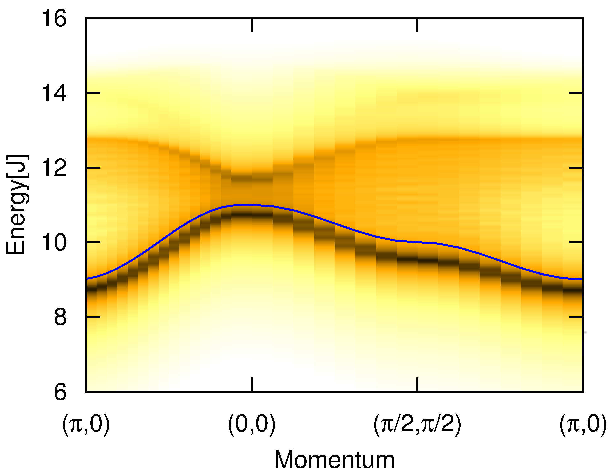}\llap{
  \parbox[b]{2.45in}{(b)\\\rule{0ex}{1.0in}
  }}\label{2b}
}
\subfigure{
\includegraphics[width=0.465\linewidth]{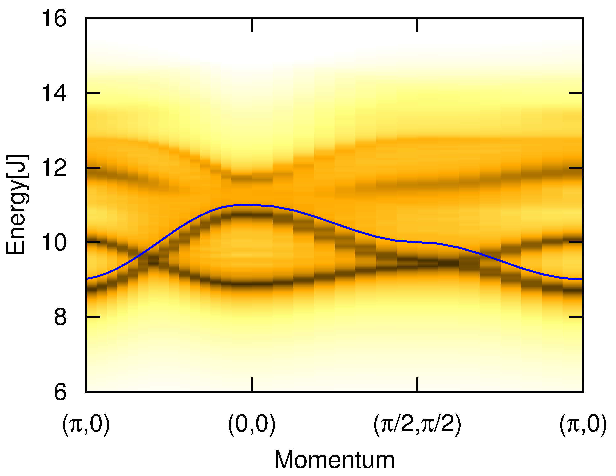}\llap{
  \parbox[b]{2.45in}{(c)\\\rule{0ex}{1.0in}
  }}\label{2c}
  }
\subfigure{
\includegraphics[width=0.465\linewidth]{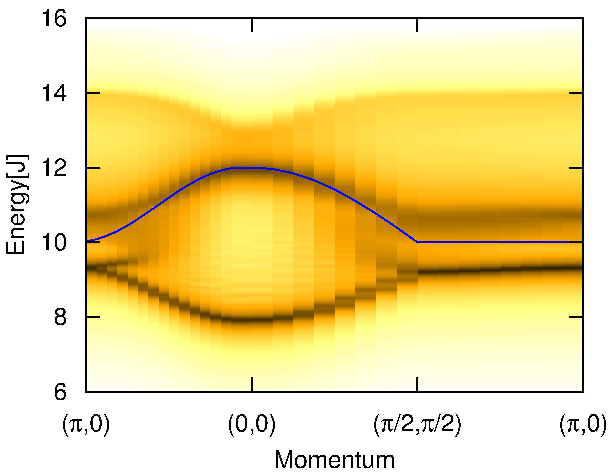}\llap{
  \parbox[b]{2.45in}{(d)\\\rule{0ex}{1.0in}
  }}\label{2d}
}
\caption{Toy model illustrating the interplay between the polaronic hopping and the various types of the free excitonic hopping:
(a) no free hopping, (b) diagonal free hopping of next-nearest-neighbor type,
(c) off-diagonal free hopping of next-nearest-neighbor type,
(d) off-diagonal free hopping of nearest-neighbor type; see text for further details.
The solid line denotes the dispersion arising from pure free hopping (i.e. the polaronic hopping is not included).\label{2}}
\end{figure}

In order to better understand the interplay of polaronic and free hopping processes in the Jahn-Teller and superexchange models,
we introduced a toy model which is based on the above-written polaronic form of the Jahn-Teller model -- though with modified
polaronic and free hopping couplings in the following way:

First of all, we assume that the longer range exchange between the $j_{\textrm{eff}}=1/2$ magnons vanishes, i.e. $J_2=J_3=0$.
Secondly, we assume a diagonal form of the matrix describing the polaronic hopping: $\hat{m}^{\rm JT} \rightarrow  \mathbb{I}$.
Next, we consider four different forms of the free hopping processes:

In the first place, we put $\hat{E}^{\rm JT}_{{\bf k}} \rightarrow 0$ -- the corresponding spectral function, calculated using SCBA (see main text of the paper), is shown in Fig.~\ref{2a}. 
It is interesting to note that adding a next-nearest-neighbor free excitonic hopping with only diagonal elements between different flavors of the excitons, i.e. substituting
$ \hat{E}^{\rm JT}_{{\bf k}} \rightarrow  z V   \mathbb{I} \cdot \left|\gamma_{2 \bf k} \right|$ (where $\gamma_{2 \bf k}=\cos k_x \cos k_y$), 
does not change the generic features of the spectral function a lot, see Fig.~\ref{2b}. Since the latter case qualitatively resembles the superexchange
model for the excitonic hopping, as discussed in Ref.~\cite{Naturecom2014Maria} and in the main text of the paper, this means that 
within the superexchange model the polaronic and the free hopping are responsible for the qualitatively similar features in the spectral function.
This is because, in the superexchange case both the polaronic and the free hopping allow for an effectively next-nearest-neighbor type of the excitonic dispersion.

In the next step, we switch off the diagonal terms in the matrix describing the free excitonic hopping and instead introduce the off-diagonal free hopping -- in order to mimic the Jahn-Teller model.
More precisely, we substitute $\hat{E}^{\rm JT}_{{\bf k}} \rightarrow  z V   \mathbb{A} \cdot \left|\gamma_{2 \bf k} \right|$,
where matrix $\mathbb{A}$ has the form:
 \begin{align}
 \mathbb{A}=\begin{pmatrix}
 0 & 0 & 1 & 0 \\
 0 & 0 & 0 & 1 \\
 1 & 0 & 0 & 0 \\
 0 & 1 & 0 & 0
 \end{pmatrix}
 \end{align}
As one can easily see in Fig.~\ref{2c}, involving the non-diagonal elements in the free hopping matrix instead of the diagonal ones
drastically changes the spectrum -- in particular, each of the two dispersive branches splits now into two branches.
Finally, the spectrum in Fig.~\ref{2d} is calculated for a toy model which also has the off-diagonal free hopping elements in the matrix -- however, instead of the next-nearest-neighbor hopping 
it includes solely the nearest-neighbor hopping [i.e. we substitute $\hat{E}^{\rm JT}_{{\bf k}} \rightarrow  z V \mathbb{A} \cdot \left|\gamma_{\bf k} \right|$].
We note that the latter case of the toy model is the closest (out of all four toy models discussed) to the considered in the main text Jahn-Teller model.

One can see that the spectra in Figs.~\ref{2c} and~\ref{2d} have slightly more in common than the spectra in Figs.~\ref{2b} and~\ref{2c}.
This means that the presence of the off-diagonal hopping elements in the free excitonic hopping plays an even more important role in the propagation
of the exciton, than the type of the free excitonic hopping dispersion (i.e. whether it is of the nearest- or next-nearest-neighbor character).

Altogether, we have shown that the particular features found in the excitonic spectrum of the Jahn-Teller model, which make it so different with respect to the superexchange model, originate 
from: (i) the nearest-neighbor-character of the free hopping that is always present in the Jahn-Teller Hamiltonian and has no analog in the superexchange model, 
and (ii) the off-diagonal elements in the free hopping matrix -- which is also absent in the superexchange case.

\bibliographystyle{prsty-etal}

\end{document}